# Thermochromic properties of 3C-, 6H- and 4H-SiC polytypes up to 500°C

G. Ferro[1,a], D. Carole[1,b], D. Chaussende[2,c]

[1]Laboratoire des Multimatériaux et Interfaces, Université de Lyon, 6 rue Victor Grignard, 69622 Villeurbanne, France

[2]Univ. Grenoble Alpes, CNRS, Grenoble INP, SIMAP, 38000 Grenoble, France

[a]gabriel.ferro@univ-lyon1.fr, [b]davy.carole@univ-lyon1.fr, [c]didier.chaussende@grenoble-inp.fr

**Keywords:** Thermochromic, Color, Temperature, Optical absorption, Polytypes

**Abstract.** The thermochromic properties (color change with temperature) of n type doped SiC wafers of different polytypes (3C, 4H and 6H) have been investigated up to 500°C under air. It was found that 3C-SiC color passes from bright yellow at room temperature to deep orange at 500°C leading to a color contrast (ΔE) as high as 64. The hexagonal polytypes undergo also a color change upon heating but far less pronounced, with ΔE values <20. All these semiconductors undergo band gap shrinkage upon heating which effect largely participated to the observed color change. This effect is very sensitive for 3C polytype since its bandgap is already in the visible energy range at room temperature. The thermochromicity of 3C-SiC was found to be reversible thanks to its thermal stability and its resistance towards oxidation.

**Introduction**

Reversible thermochromic materials undergo a color change upon temperature variation, which color is fully restored upon coming back to the initial temperature [1]. Several applications have been reported for temperature below 100°C, such as smart windows [2], clothing [3] or even ink [4]. When targeting higher temperatures (up to 500°C for instance), commercially available colored oxide semiconductors (e.g. $BiVO_4$ [5], $Bi_2O_3$ or $WO_3$ [6]) with bandgap in the visible range are preferred due to their supposed high thermal stability under air. This is not exactly true since these transition metal oxides can lose O atoms at high temperature generating O vacancies which affect the reversibility of their thermochromism [7,8]. On the other hand, the thermochromic properties of oxidation resistant non-oxide semiconductors have been poorly investigated so far. 3C-SiC, whose bandgap of 2.39 eV at RT, is known to be yellow. And according to the early work of W.J. Choyke, its linear bandgap shrinkage with increasing temperature (down to ~2.2 eV at 500°C) [9] should lead to detectable color change by naked eye. It is surprising that, despite all the works done so far on 3C-SiC, no paper dealing with the possible thermochromic properties of this SiC polytype can be found in the literature. This is the main goal of the present study. Additionally, since 6H and 4H polytypes of SiC have distinguishable color at RT (due to impurity doping), they will be also tested for thermochromicity.

**Experimental**

The materials used for this study are pieces of single crystalline wafers (n-type doped) of each polytype: from commercially available source for 4H and 6H and from former Hoya company for 3C. A homemade set-up was used for both qualitative and quantitative color estimation in combination with heating (see ref [6] for more details). Briefly, the heating stage (thermocouple-controlled) is positioned inside a closed black-box which can bed uniformly illuminated from the top. A camera placed above an aperture on the top side can take pictures of the samples during heating. The SiC wafers were placed on a dull-white boron nitride (BN) holder in order to ensure good thermal conduction while avoiding i) chemical reactivity with the heating metallic stage and ii) unwanted reflection of the light from the stagewhich could induce color estimation errors. Some graphite powder and a BaSO$_4$ pellet are placed beside heating stage (and thus unheated), respectively for fixing the black and white levels of the images. A heating cycle involveda succession of monitored heating ramps (10°C/min) followed by temperature plateaux (3 min duration) at 100, 200, 300, 400 and 500°C.The pictures were taken after 2 min of temperature stabilization at each plateau. Additional pictures were taken at room temperature (RT) before heating and after cooling back to RT (back RT) for reversibility check.

Each picture was analyzed using GIMP software. The procedure consisted in adjusting first the color tunes according to the black and white references of the photos (graphite powder and BaSO$_4$). Then the CIE-Lab colorimetric parameters (L*a*b*) of each sample were extracted from these images, considering an average area of 20x20 pixels taken at the center of the samples. The color change with temperature can be better assessed by calculating the color contrast (ΔE) versus temperature using the following formula:

$$\Delta E = \sqrt{(L_T^* - L_{RT}^*)^2 + (a_T^* - a_{RT}^*)^2 + (b_T^* - b_{RT}^*)^2}$$

with L$^*_T$, a$^*_T$ and b$^*_T$ the CIE-Lab parameters at the measured temperature T and L$^*_{RT}$, a$^*_{RT}$ and b$^*_{RT}$ the CIE-Lab parameters of the same sample but at RT.

**Results and discussion**

Since the perceived colors are the sum of absorptions and reflections occurring at different wavelength of the visible spectral region, the samples were characterized by UV-Vis absorption spectroscopy in order to discuss on the origin of their color. Results are shown on Figure 1. One can see that 3C-SiC is the only polytype showing typical absorption expected for a pure semiconductor, with an abrupt threshold corresponding to its optical gap. The hexagonal polytypes both display broad absorption bands below the gap while their mean absorption level in the visible region is rather high as compared to 3C. These spectra for 6H and 4H are very similar to the ones reported in [10] for N doped samples. Altogether, it gives at RT a bright yellow color to 3C, a greyish green for 6H and a greenish brown for 4H (see Figure 2a).

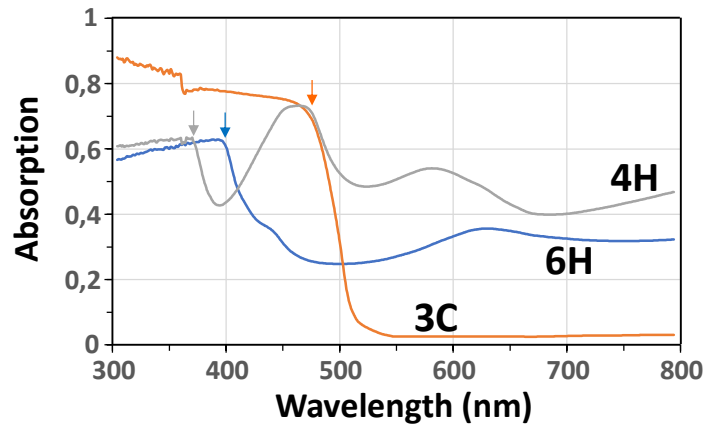

**Figure 1**. UV-Vis absorbance spectra recorded at RT on each polytype of this study. The colored arrows correspond to the respective bandgap absorption thresholds.

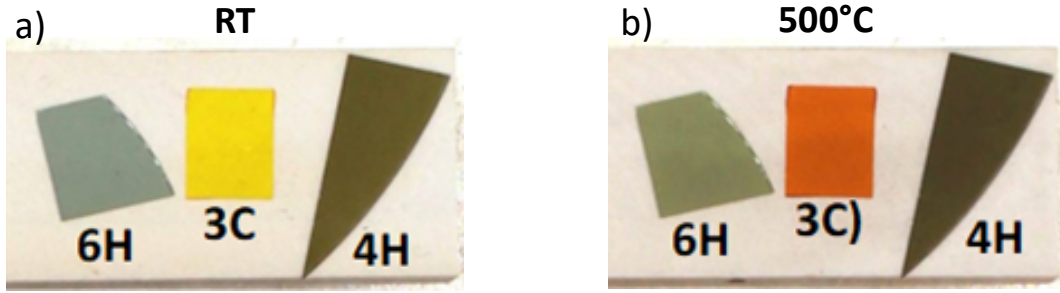

**Figure 2.** Pictures of the SiC wafers taken at a) RT and b) 500°C.

After heating at 500°C, the color of all wafers has changed (Fig. 2b). This is rather faint for the hexagonal polytypes (4H gets darker brown while 6H gets yellowish green) but remarkable for 3C (it gets deep orange). 3C-SiC is thus clearly a thermochromic material. To quantitatively evaluate this thermochromism, the evolution of ΔE with temperature is drawn in Fig. 3. Details of the colorimetric parameters extracted from each picture are given in Table 1.

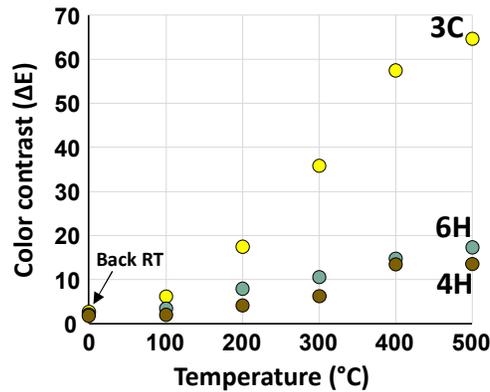

**Figure 3**. Evolution of the color contrast as a function of temperature for each polytype.

**Table I.** Evolution of the colorimetric parameters as a function of temperature for each polytype.

| Temperature (°C) | 3C | | | 4H | | | 6H | | |
|---|---|---|---|---|---|---|---|---|---|
| | L* | a* | b* | L* | a* | b* | L* | a* | b* |
| RT | 87.4 | -3.5 | 84.2 | 37 | 0.9 | 27.7 | 61.7 | -7.3 | 7.8 |
| 100 | 85 | 2.1 | 82.6 | 36 | -0.9 | 27.3 | 62.2 | -7.9 | 10.5 |
| 200 | 80.5 | 12 | 79.3 | 36.4 | -3,4 | 27,1 | 62.8 | -10.2 | 15.1 |
| 300 | 70 | 25.1 | 70.5 | 33.2 | -0.8 | 22.1 | 60.7 | -7.2 | 18.3 |
| 400 | 55.8 | 37.6 | 59.9 | 27.6 | 1.9 | 17.6 | 52.5 | -3.9 | 18.2 |
| 500 | 55.1 | 46.9 | 60.6 | 27.9 | 0.9 | 17.5 | 56.4 | -6.1 | 24 |
| Back RT | 87.5 | -4.8 | 80,6 | 36,5 | 0.9 | 27 | 61 | 6.9 | 8.1 |

Upon cooling back to RT, the ΔE values are below 3 for all polytypes which means that the difference is hardly noticed by human eye [11]. This confirms the reversibility of the observed thermochromism. This was an expected result since SiC material is very inert and no significant change is anticipated upon heating to 500°C under air except a very small surface oxidation in the nm range [12]. According to the absorbance spectra shown in Fig. 1, the thermochromism of 3C-SiC is the simplest to explain. Indeed, as it shows no optical absorption at RT except the one related to its bandgap, the observed color change upon heating for 3C-SiC is most probably due to its bandgap shrinkage. For the other polytypes, their color at RT does not come from their respective bandgapbut rather from the N impurity-related optical absorption below the gap [10] (comparatively, it is known that high purity (semi-insulating) 4H and 6H crystals are transparent and colorless).

As a matter of fact, the moderate thermochromism of N doped 4H and 6H polytypes is probably due to the low sensitivity to temperature of these N-related absorptions. Their bandgap shrinkage with increasing temperature should contribute negligibly to their color since these gaps should reach ~3.07 and ~2.83 eV at 500°C respectively for 4H and 6H (according to [9]). Note that 2.83 eV bandgap for 6H could be low enough to bring a very faint yellow tune which is indeed observed at 500°C, as discussed earlier. It could thus explain why 6H has a slightly higher thermochromism than 4H.

In order to evaluate the interest 3C-SiC as a thermochromic material, let us compare with other already known and commercially available compounds. This is summarized in Table II.

**Table II.** Comparison of the color contrast $\Delta E$ of 3C-SiC with different commercially available compounds (powders) having thermochromic property.

| Compound | Temperature (°C) | $\Delta E$ | Reference |
| --- | --- | --- | --- |
| $Bi_2O_3$ | 500 | 65 | [6] |
| $BiVO_4$ | 280 | 61 | [5] |
| $WO_3$ | 500 | 53 | [5] |
| $Fe_2O_3$ | 500 | 36 | [6] |
| 3C-SiC | 500 | 64 | This work |

3C-SiC is obviously among the best thermochromic compounds, though $BiVO_4$ thermochromicity is probably the strongest since its $\Delta E$ value was obtained at 280°C only (one expects this color contrast to increase significantly when heating up to 500°C). However, as mentioned earlier, the thermochromic reversibility of this oxide is questionable due to its tendency to easily lose O atoms upon heating, under inert gas or even under air [7,8]. This is also the case for $Bi_2O_3$, but to a lower extent. As a matter of fact, the high chemical inertness of 3C-SiC associated with the good values of $\Delta E$ make it a promising material for thermochromic application at high temperature under air or harsh environment. The next step of this study would be to make some cycling tests (10 to 20 times) for confirming this reversibility. Testing some powder (which is the form commonly used in thermochromic applications) made of high purity 3C-SiC would be also of interest for checking the difference (if any) with the single crystal.

## Summary

We have evaluated the thermochromic properties of n-type doped 3C-, 6H- and 4H- polytypes of SiC up to 500°C. 3C-SiC shows by far the strongest color change (from yellow to deep orange) with $\Delta E$ as high as 64. Its thermochromicity seems to be fully reversible (to be confirmed by cycling tests).

## Acknowledgements

The authors would like to acknowledge Dr Etienne BUSTARRET and Dr Valerio OLEVANO from Néel Institute (UPR 2940, Grenoble) for their contribution through fruitful discussions.